# Understanding the Systems Biology of Pathogen Virulence Using Semantic Methodologies

David Rhee[*], Kevin Shieh[*], Julie Sullivan[†], Gos Micklem[†], Kami Kim[‡] and Aaron Golden[*][§]
[*]Department of Genetics
Albert Einstein College of Medicine, Bronx NY 10461
Email: david.rhee@einstein.yu.edu, kevin.shieh@med.einstein.yu.edu
[†]Department of Genetics
University of Cambridge, Cambridge CB2 3EH, U.K.
Email: jms237@cam.ac.uk, g.micklem@gen.cam.ac.uk
[‡]Departments of Medicine, Pathology, Microbiology & Immunology
Albert Einstein College of Medicine, Bronx NY 10461
Email: kami.kim@einstein.yu.edu
[§]Department of Mathematical Sciences
Yeshiva University, New York NY 10033
Email: aaron.golden@einstein.yu.edu

*Abstract*—Systems biology approaches to the integrative study of cells, organs and organisms offer the best means of understanding in a holistic manner the diversity of molecular assays that can be now be implemented in a high throughput manner. Such assays can sample the genome, epigenome, proteome, metabolome and microbiome contemporaneously, allowing us for the first time to perform a complete analysis of physiological activity. The central problem remains empowering the scientific community to actually implement such an integration, across seemingly diverse data types and measurements. One promising solution is to apply semantic techniques on a self-consistent and implicitly correct ontological representation of these data types. In this paper we describe how we have applied one such solution, based around the InterMine data warehouse platform which uses as its basis the Sequence Ontology, to facilitate a systems biology analysis of virulence in the apicomplexan pathogen *Toxoplasma gondii*, a common parasite that infects up to half the worlds population, with acute pathogenic risks for immuno-compromised individuals or pregnant mothers. Our solution, which we named 'toxoMine', has provided both a platform for our collaborators to perform such integrative analyses and also opportunities for such cyberinfrastructure to be further developed, particularly to take advantage of possible semantic similarities of value to knowledge discovery in the Omics enterprise. We discuss these opportunities in the context of further enhancing the capabilities of this powerful integrative platform.

*Keywords*-data warehouse; genomics; proteomics; epigenomics; metabolomics; sequence ontology; integrative analysis; systems biology; intermine; pathogen; toxoplasma gondii

## I. INTRODUCTION

### A. Semantic Computing & Systems Biology

The technological revolution currently under way in the life sciences is not only transforming our knowledge discovery potential in these disciplines, but is also starting to make real change in human healthcare, and as a consequence, society as a whole. The use of 'desktop' massively parallel sequencing platforms (to study the properties and content of nucleic acids), chromatography/mass spectrometry (similarly for the proteins and metabolites that make cells 'work'), and the efficient management of experimental meta-data using various Laboratory Information Management Systems (LIMS) schema offer a powerful means to fully characterize the physiology of a given cell, organ, indeed an entire organism, for the first time.

The promise of leveraging such empirical systems biology methodologies offer a means to construct realistic and empirically based models of such physiological activities. However, such characterization is profoundly constrained by our ability to assimilate in a meaningful way these same data products in such a way as to permit the construction of viable systems biology models. Semantic technologies offer perhaps the only means we can ever hope to synthesize and meaningfully explore the diversity of digitized data products generated from the numerous molecular assays that have been automated in the past few decades.

### B. Data Warehousing & Sequence Ontology

An obvious first step in integration of multiple and diverse data types - such as RNA-seq, ChIP-seq, SNP array, proteomic & metabolomic data - is to federate each data repository using a data warehouse. In the best traditions of computer science, the algorithms that define how the federation data structure will be implemented also provide the means by which the data can be interrogated in its entirety. However, when one considers the significant diversity in data product types used in the digitized genomics space, one needs to define an appropriate and correct ontology within which to not only make these data instances semantically

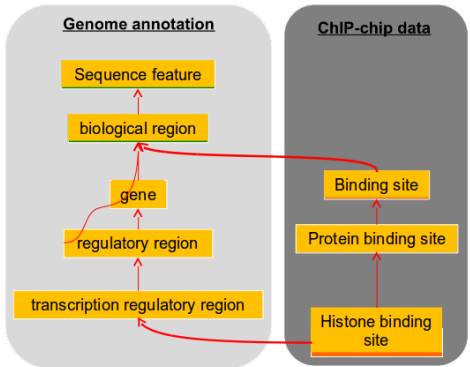

Figure 1. Visual representation showing how an experiment data instance - in this case from a ChIP-chip microarray experiment - is mapped into genome annotation space using the Sequence Ontology

relevant with respect to each other, but also to facilitate exploration of relationships embedded in the experimental data space. Here we can take advantage of the fundamental nature of any organism's genome, in that an ontology, based on community accepted definitions of genomic sequence features that are used to annotate such genomes, can be defined. This framework, known as the Sequence Ontology [1], has been developed over many years in the genomics community, and takes advantage of the fact that $any$ biological process or context may be expressed using this fundamental schema. In this way, one can efficiently and self-consistently integrate fundamental genomic, epigenomic, proteomic and metabolomic datasets along with higher level pathway, gene ontology and experimental metadata entities within the same data model space - and consequently explore the integrated data space, using the embedded semantic relationships to facilitate such analyses (Figure 1).

*C. The InterMine Data Warehouse Platform*

To date probably the most robust and widely used data warehouse platforms based on the Sequence Ontology (SO) is InterMine [2]. At its core is the implementation of a highly flexible data model, based on the SO schema, that allows the incorporation of a wide range of existing biological data types $and$ the ability to easily add new data types, simply by modifying associated XML files. The system architecture is centralized around the ObjectStore, a Java based custom object/relational mapping system optimized for read-only access to a PostgreSQL database hosting the collated biological data. The query vocabulary is based around standard set theory and Boolean operators, which provide a means to interrogate the data spaced encapsulated by the underlying data model. Such queries are interpreted by the ObjectStore into SQL and results returned from the PostgreSQL database, and this can be mediated through RESTful web services (currently capable for Python, Perl, Ruby, Java and Javascript) and directly via a web GUI. InterMine also supports the development of embedded 'widgets' whereby visualization of the SQL results is possible, showing tables, visual charts and enrichment analysis plots. One particularly attractive aspect of the InterMine system is its innate interoperability - as the schema is based around SO, and because all living organisms possess genomes, it is possible to query across different InterMine instances for common homologs, enabling comparative analysis. Several InterMine instances are operational at present for several organisms including FlyMine [3], YeastMine [4], MouseMine[5], MetabolicMine [6], MitoMiner [7], INDIGO [8], and as we describe here, toxoMine [9]. Specific variants also exist for drug discovery (TargetMine [10]) and to support the modENCODE project (modMine [11]). The latter is a particularly pertinent example as the aim of the modENCODE project [12] was to identify all common genomic functional elements in the model organisms *Drosophila melanogaster* and *Caenorhabditis elegans* by integrating high-throughput datasets generated for each.

## II. TOXOPLASMOSIS & HUMAN HEALTH

The pathogen *Toxoplasma gondii* (*T. gondii*) causes the disease toxoplasmosis and is an obligate intracellular parasite. It is classified as a Category B Biodefense agent and has been estimated to infect up to half of the world's population. Its life cycle involves reproduction uniquely within the cat gastrointestinal tract, the subsequent distribution of oocysts ("spores") in cat fecal material and the ingestion of these oocysts by almost any warm-blooded mammals. Initial infection is typically suppressed by the immune system, with the parasite remaining in a latent state typically in cysts that form in the nervous and muscle tissue.
Whilst many infected individuals are asymptomatic, people with compromised immune systems (e.g. patients undergoing cancer therapy, who have received an organ transplant or are living with HIV) and pregnant women are particularly vulnerable to the effects of this parasite in its active state, where the effect on new borns can be severe. A member of the Apicomplexan parasitic group (which includes the genus $Plasmodium$, the cause of malaria), *T. gondii* regularly monitors its environment within the infected host, and evidence suggests that epigenetic regulation, changes in gene expression and subsequent activation/deactivation of gene networks play a role in the regulation of its pathogenic potential and virulence traits as it adapts to such changes [13][14]. Whilst the dominant strains of *T. gondii* have been fully sequenced, many aspects of its genome and associated functional properties remain to be elucidated, and so several initiatives are underway to try and discern the methods by which this parasite modulates its interaction especially with the human host.

*A. A Systems Biology Approach to the Model Apicomplexan Toxoplasma gondii*

The NIH-funded 'A Systems Biology Approach to the model Apicomplexan Toxoplasma Gondii' (led by K. Kim at the Albert Einstein College of Medicine) collaborative project has the specific goal of integrating ChIP-chip, ChIP-seq, RNA-seq, proteomics data in such a way as to directly employ systems biology methodologies to understand this parasite's regulatory processes. Working with this group it became clear to us that the InterMine platform would provide an ideal means of representing the resulting dataspace, and in particular of facilitating its exploration consistent with systems biology protocols. In the next section we describe the resulting operational instance, toxoMine [9].

*B. Design & Current Implementation of toxoMine*

We used the basic InterMine system as the foundation for toxoMine. The latest genome build for *T. gondii* and associated annotation datasets were obtained from the ToxoDB public resource [15] (Release 11 for the three most widely used strains ME49, GT1 & VEG). Protein and protein domain data were incorporated from Uniprot [16] and InterPro [17], homolog data from the OrthoMCL database [18] and Gene Ontology data from the Gene Ontology consortium archives [19]. We also uploaded all relevant publication data on *T. gondii* from the PubMed bibliographic repository. This formed the publically available data space. We developed a custom parser to support the import/export of locally derived experimental data in a local 'staging' database into the toxoMine system, combined with a defined data model that captured our experimental data model within the SO schema. In the post-processing stage, the object entities within toxoMine's database are linked via the underlying data model, yielding an attribute index which is critical to effective query implementation. toxoMine uses two Linux servers interchangeably, each with 12-core Intel processors, 96 GB RAM and 4 TB RAID10 storage as both build and production systems. The web portal is supported by Apache Tomcat. toxoMine is publically accessible at http://toxoMine.org and is fully functional [9] (Figure 2). Currently the system is operational and actively undergoing population and actively used as part of the 'A Systems Biology Approach to the model Apicomplexan Toxoplasma Gondii' project (K. Kim, Einstein and collaborators), thus providing excellent evaluation feedback for its early stage implementation. Further dissemination will take place when this work is presented at the next International Congress on Toxoplasmosis & *T. gondii* Biology, at which point we will have been able to fully integrate feedback on improving usability of this resource.

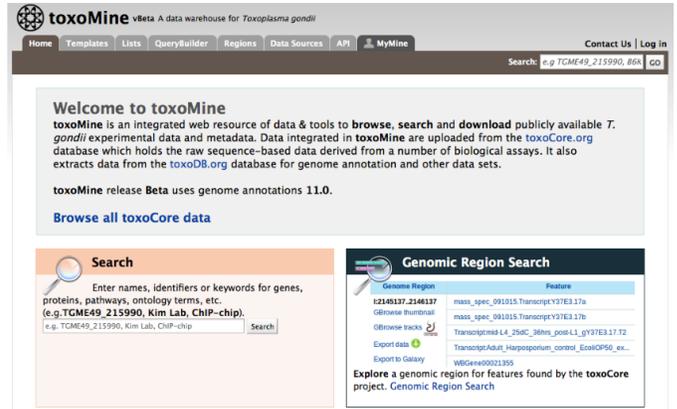

Figure 2. The web portal to the toxoMine data warehouse as currently operational

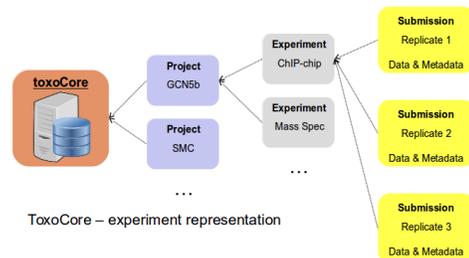

Figure 3. Schematic showing hierarchical manner in which experimental datasets will be integrated into the toxoCore system

*C. Current Activities: Enabling 'hot pluggable' data entry - toxoCore*

As technology continues to improve and prices to drop, small and medium sized research groups are implementing a range of high throughput molecular assays and undertaking projects at a scale that previously would only have been possible within consortia or at genome centers. Thus the demand for integrative solutions for data analysis and dissemination is increasing. Our experience to date makes it clear that significant time is usually required to prepare the raw data in such a way that it, and its associated metadata, are correctly formatted for inclusion into the toxoMine database.

With this in mind we have been working towards this future of systems biology research, by optimizing the process for carrying out regular data updates within InterMine. Our solution, known as 'toxoCore', is a database system separate from that of the toxoMine, and is designed exclusively to upload, store and manage in a controlled manner high-throughput experimental data and metadata (Figure 3). In the first instance we will implement a supervised update procedure involving the 'toxoMine' administrator, but we envisage an alternative automated scheduled update mechanism to be developed in due course. We are working on adapting the ISA-TAB standard [20] to capture all meaningful experimen-tal metadata and permit its subsequent integration into the toxoCore database. Critical to this process is a transparent and efficient means of uploading such experimental datasets, and we are in the process of finalizing both GUI and command line interfaces to this end. Coordination between both the separate toxoCore server and the main toxoMine server can be configured at the systems level, permitting regular updates to the toxoMine database at suitable times to minimize user disruption (Figure 4). The fact that the InterMine system is open source means that other InterMine instances will be able to adopt this and other system-specific innovations to other systems biology applications, making use of the more generic components such as 'Project', 'Experiment' and 'Submissions'.

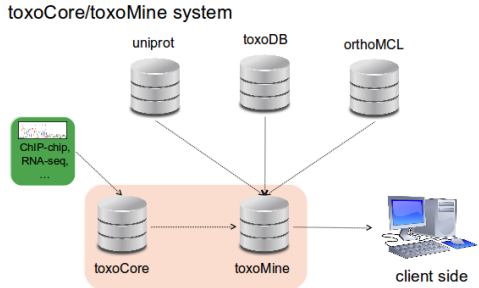

Figure 4. Schematic showing toxoMine system as implemented. Public repositories provide the core data components for the system's PostgreSQL DB, with locally derived data products periodically uploaded to the toxoCore repository prior to integration into the same core DB.

toxoCore is in the development phase and we hope to have it operational in 2016. Whilst this additional component will significantly improve the ability of our domain experts actively working on *T. gondii* to add new data products to toxoMine, it also offers a new paradigm for InterMine deployment more generally, particularly for those small/medium sized enterprises.

### D. Future Opportunities: Enhancing user interactivity

As stated, the current InterMine release permits the construction and deployment of queries to the ObjectStore either directly using a query web-page, or via several APIs that permit RESTful interaction with the system, currently via Python, Perl, Ruby, Java and Javascript. However the formulation of queries can be a laborious process, as the exact syntax and semantic structure of the queries requires the adoption of set rules and pre-defined relationships determined by the underlying data model. Certainly, through the web front end, the combination of predicate-object units in a logical cascade is cumbersome to determine using the drop-down menus supplied, and construction of a query via an API, whilst guaranteeing efficient implementation and return, is still a tedious task. The InterMine team invoked the idea of 'templates', in effect atomic or more complex fundamental query types (i.e. involving several query operators in a logical workflow) that could be re-used as appropriate, but this would tend to limit most researchers use of these systems to these most basic query operations, human nature being what it is. Ideally one needs to overhaul the query experience, in a way that permits users who aren't sophisticated bioinformaticians (i.e. a typical biomedical scientist) to directly and effectively interact with their data. The fact that we have an underlying data model linking the data objects (and their associated functional and semantic properties) means that we could, in theory construct more graphically oriented ways of constructing query 'strings' - or workflows. To that end we are currently assessing the capability of existing modules that permit the graphical construction and deployment of such workflows, such as Taverna's Workbench [21], in the InterMine context, as we believe such an implementation would dramatically enhance user interactivity with InterMine systems - either via a GUI or in the conversion of 'correct' graphically constructed workflow queries into say Python for command-line deployment.

We are also exploring an alternative means of implementing a natural language based approach, such as the TR Discover package [22], which provides an efficient and optimized means for a user to formulate valid natural language question(s) based on the entities & relationships already present and underpinning the data model. In the shorter term we are also working on the development of a basic R API, the most widely used and dominant statistical processing environment amongst the bioinformatics community [23], to facilitate direct interaction with the InterMine ObjectStore.

### E. Future Opportunities: Consolidating interoperability with other InterMine instances

Perhaps the most appealing aspect of the InterMine project is its basis in the idea of the Sequence Ontology - just like every living organism, every InterMine instance shares the same genomic concepts and foundations, and consequently, interoperability between instances is greatly facilitated. The InterMOD consortium [24] has been working to demonstrate such interoperability across a number of InterMine instances for the 'model' organisms widely studied in the life sciences. The consortium uses sequence homology as a basis for mapping between analogous genes in the different organisms. Therefore toxoMine, by incorporating homology data from the OrthoMCL database [18], can currently send cross-species queries to other InterMOD compliant systems to determine if genes identified from an analysis in one organism have certain properties or characteristics as a group, or part of a group, in another organism. This follows the implicit logic of the genome itself being the common reference frame to all organisms, and that even if divergent evolution from common ancestors has altered the content of many genes, their ultimate function is very often preserved. However a real opportunity exists in leveraging the rest of the unified schema that forms the basis for InterMine's data model, through the application of semantic similarity techniques driven on other contextually relevant queries, beyond simple Gene Ontology relationships. For example, the use of such a methodology to discern what are likely batch effects would be a tremendously powerful means to self-consistently validate the results of any integrative analysis.

*F. Future Opportunities: Optimizing Data Model Construction*

Data modelling is an important and necessarily difficult part of the process of realizing a new databases. The Sequence Ontology-based core data model of InterMine is often a good starting point and from this essentially all of the rest of the InterMine infrastructure (programming interfaces, classes, database infrastructure, web interface) is generated automatically. Nevertheless there can be significant effort required to appropriately model new data sources and editing and updating the the InterMine data model could be made easier. This is important as it is absolutely essential to modify the underlying data model whenever new molecular assays are implemented or new experimental paradigms adopted. Despite the diversity of molecular assays in current use, there are a core set of fundamental methods (e.g. ChIP-seq, RNA-seq) and this core set can form the foundational basis for any SO-based genomics schema. This basic configuration can be codified as a core UML model and we are considering re-factoring of tools such as the National Cancer Institutes's caCORE SDK [25] to provide a more versatile and efficient means of interacting with InterMine's data model representation.

## III. CONCLUSION

Charged with the goal of developing an informatics environment that would allow our colleagues to explore multiple genomics, epigenomics and proteomics datasets obtained from experiments studying virulence in the parasite *Toxoplasma gondii*, we commissioned a new instance of the integrative data warehouse system InterMine. The resulting platform, toxoMine, has permitted our colleagues to perform systems biology based studies of the regulatory processes of this pathogen and in so doing, significantly enhance their scientific and biomedically relevant research. Commissioning toxoMine also provided us with a means to assess the strengths and weaknesses of the InterMine system particularly for those members of the potential user community who simply do not possess the resources common to large consortia, which have to date dominated the InterMine user community. We identified areas for improvement in (i) optimizing the 'upload' of new experimental data to the existing toxoMine system (ii) improving the user experience as regards the construction and deployment of queries to the data warehouse (iii) exploring the use of semantic similarity methods to enhance exploration of the full dataspace, both within a given InterMine instance and between a set of instances each representing a unique organism or biochemical context, and (iv) developing more efficient means of instantiating and curating the basic 'data model' that forms the core of any InterMine instance. Taken together these work areas offer a blueprint by which the InterMine data warehouse concept could be significantly enhanced, ultimately providing a more turn-key integrative solution for biomedical and life scientists hoping to fully profit from the fruits of their labors using today's high throughput technologies. In this regard, working with our colleagues on ongoing study of the *Toxoplasma gondii* parasite provides an excellent test-bed environment to baseline these efforts, both currently under way and planned.

## ACKNOWLEDGMENT

The authors acknowledge support from the National Institute of Allergy and Infectious Diseases (Grant Number: RC4AI092801 to Dr. K. Kim). This work was also supported by the Wellcome Trust (Grant number 099133 to Dr. G. Micklem) and National Human Genome Research Institute (Grant Number: R01HG004834 to Dr. G. Micklem). The contents of this paper are solely the responsibility of the authors and do not necessarily represent the official views of the funding bodies.